\input harvmac.tex
\vskip 2in
\Title{\vbox{\baselineskip12pt
\hbox to \hsize{\hfill}
\hbox to \hsize{\hfill HIP-2002-05/TH}}}
{\vbox{\centerline{On Noncommutativity in
String Theory and D-Branes}
\vskip 0.3in
{\vbox{\centerline{}}}}}
\centerline{Masud Chaichian{$^\dagger$}\footnote{}
{}}
\centerline{and}
\centerline{Dimitri Polyakov\footnote{$^\dagger$}
{Masud.Chaichian@helsinki.fi, Dimitri.Polyakov@helsinki.fi}}
\medskip
\centerline{\it High Energy Physics Division}
\centerline{\it Department of Physical Sciences}
\centerline{\it University of Helsinki}
\centerline{\it and}
\centerline{\it Helsinki Institute of Physics}
\centerline{\it P.O. Box 64, FIN-00014  Helsinki, Finland}
\vskip .2in
\centerline {\bf Abstract}

String theory in a constant B-field exhibits noncommutative structure of
space-time.
By considering the B-field dynamical and studying its interaction with
Ramond-Ramond (RR) background we observe the breaking of the
B-field gauge symmetry in the effective action.
This effect takes place due to non-perturbative
coupling of the B-field to membrane topological charge.
As a result, the B-field is renormalized in the RR
backgrounds, making it impossible to obtain consistent
non-commutative models with constant B-field.
 We argue that the
gauge invariance
should be restored by introducing appropriate
external D-brane configuration.

\Date{February 2002}
\vfill\eject
\lref\fms{D. Friedan, E. Martinec and S. Shenker, Nucl. Phys.{
B271 (1986) 93}}
\lref\azc{
J. de Azcarraga, J. Gauntlett, J. Izquierdo
and P. Townsend Phys. Rev. Lett. 62 (1989) 2579}
\lref\pol{ J. Polchinski,
 hep-th/9510017, { Phys. Rev. Lett.75 (1995) 4724}}
\lref\seiberg{N. Seiberg and E. Witten, hep-th/9908142,{ JHEP 9909
(1999) 032}}
\lref\seib{N. Seiberg,
hep-th/0008013,
{ JHEP 0009 (2000) 003 }}
\lref\and{O. Andreev and H. Dorn, hep-th/9912070, Phys. Lett {
B476}(2000) 402}
\lref\shahin{ F. Ardalan, H. Arfaei and M. Sheikh-Jabbari, hep-th/9810072,
{JHEP 9902 (1999) 016}}
\lref\self{D. Polyakov, hep-th/0111227, to appear in Phys. Rev. D}
\lref\ler{P. Di Vecchia, M.Frau et al., hep-th/9906214,
Nucl.Phys.B565:397-426, (2000)}
\lref\lerd{M. Billo, P. Di Vecchia et
al.,hep-th/9802088,Nucl.Phys.B526:199-228,(1998)}

\centerline{\bf 1. Introduction}

Space-time noncommutativity in string theory appears as
a result of open string dynamics in a constant B-field background
~{\seiberg, \shahin}.
The case of a special interest is the noncommutativity in the presence of
D-branes.
Prior to the remarkable work of Polchinski ~{\pol} on D-branes
it had long been a profound puzzle which
objects are the actual sources of Ramond-Ramond charges.
In superstring theory charges are usually carried by emission
vertices of corresponding physical states, i.e. by BRST invariant
and non-trivial vertex operators.
For instance, the open superstring vertex operator of a photon:
\eqn\grav{\eqalign{V_{ph}(k)
=
A_m(k)c(\partial{X^m}+i(k\psi)\psi^m)e^{ikX}(z)\cr
k_m{A^m(k)}=0}}
is multiplied by the
vector gauge potential $A_m(k)$ of the photon
and therefore can be regarded as the source of
the U(1) electric charge.
In the Ramond-Ramond sector, however, the situation is different.
It is well-known that expressions for Ramond-Ramond vertex operators
at canonical picture are given by:
\eqn\grav{\eqalign{V_{RR}(k)={1\over{p!}}{\gamma^{m_1...m_p}_{\alpha\beta}}
F_{m_1...m_p}(k)
{c\bar{c}}e^{-{1\over2}\phi-{1\over2}{\bar\phi}}
\Sigma_\alpha{\bar{\Sigma}}_\beta{e^{ikX}}(z,\bar{z})}}
where $\gamma^{m_1...m_p}_{\alpha\beta}$ is antisymmetrized
product of 10d gamma-matrices, $\Sigma$, $\bar{\Sigma}$
are spin operators for matter fields, $\phi$ is bosonized
superconformal ghost and the rank $p$ of the RR field strength $F$
may be odd or even, depending on the type of superstring theory
we consider - that is, type IIA or type IIB.

The standard bozonization formulae ~{\fms} for reparametrization
and superconformal ghosts $b,c,\beta$ and $\gamma$ are given by:

\eqn\grav{\eqalign{
c(w)=e^\sigma(w),
b(w)=e^{-\sigma}(w),
\gamma(w)=e^{\phi-\chi}(w),
\beta(w)=e^{\chi-\phi}\partial\chi(w)\cr
<\sigma(z)\sigma(w)>=<\chi(z)\chi(w)>=log(z-w)\cr
<\phi(z)\phi(w)>=-log(z-w)}}

The crucial point here is that the BRST invariance of the operators (2)
 requires that they couple to RR field strength rather
than RR gauge potential. BRST invariance condition is then equivalent
to the Maxwell's equations for the RR field strength $F(k)$.
For this reason, the RR vertex operators (2) cannot be considered as the sources
of the RR charges. This puzzle
has  been resolved in the crucial work by Polchinski where
it has been shown that the RR charges are carried by  the non-perturbative
solitonic objects, the D-branes. Namely, the RR gauge potentials
appear in the WZ terms of DBI actions for Dp-branes, coupling
to their worldvolume $p+1$-forms. D-branes may also be realized as
open strings with mixed Dirichlet-Neumann boundary conditions, at least
up to massless modes.
Remarkably, however, the  RR  vertex operators
carrying the RR charges still can be constructed at non-canonical
pictures.
Namely, consider the physical RR vertex operator
at picture $(-3/2,-3/2)$:
\eqn\grav{\eqalign{V_{RR}^{(-3/2,-3/2)}(k)=
{1\over{p!}}
{\gamma^{m_1...m_p}_{\alpha\beta}}
R_{m_1...m_p}(k)
{c\bar{c}}e^{-{3\over2}\phi-{3\over2}{\bar\phi}}
\Sigma_\alpha{\bar{\Sigma}}_\beta{e^{ikX}}(z,\bar{z})}}
Properties of this operator are significantly
different from those of (2) (this question
has also been considered in ~{\lerd, \ler}.
It is easy to check that in this case
 BRST invariance condition imposes no
on-shell constraints on the polarization p-form
$R_{m_1...m_p}(k)$ and for this reason it cannot be
interpreted as a RR field strength.
We shall refer to  the p-form space-time field
$R_{m_1...m_p}(k)$ as Ramond-Ramond prepotential,
for reasons which will later become clear.
To point out the physical meaning of
$R_{m_1...m_p}(k)$ let us act on it with
left and right picture-changing operators
$\Gamma$ and $\bar{\Gamma}$ and compare the result with (2).
The normal ordered expression for $\Gamma$ is given by

\eqn\grav{\eqalign{\Gamma(z)=:e^\phi(G_{matter}+G_{ghost}):(z)
={1\over2}e^\phi\psi_m\partial{X^m}(z)+
{1\over4}be^{2\phi-\chi}(\partial\sigma+\partial\chi)(z)
+c{e^\chi}\partial\chi(z)}}
and analogously for
$\bar{\Gamma}$.

Using the relevant OPE's, including
\eqn\grav{\eqalign{\psi^m(z)\Sigma_\alpha(w)\sim
(z-w)^{-{1\over2}}\gamma^m_{\alpha\gamma}\Sigma_\gamma(w)+...\cr
e^{\alpha\phi}(z)e^{\beta\phi}(w)\sim(z-w)^{-\alpha\beta}e^{(\alpha+\beta)\phi}
(w)+...\cr
\partial{X^m}(z)e^{ikX}(w)\sim(z-w)^{-1}ik^m{e^{ikX}(w)}+...\cr
\Sigma_\alpha(z)\Sigma_\beta(w)\sim{{\delta_{\alpha\beta}}\over
{(z-w)^{5\over4}}}+\sum_p{1\over{p!}}
{{\gamma^{m_1...m_p}_{\alpha\beta}
\psi_{m_1}...\psi_{m_p}(w)}\over{(z-w)^{{5\over4}-{p\over2}}}}+...
}}
we have:
\eqn\grav{\eqalign{:{\bar\Gamma}{V_{RR}^{(-3/2,-3/2)}(k)}:=
{1\over2}({\gamma^{m_1...m_pn}_{\alpha\beta}}
R_{{\lbrack}m_1...m_p}(k)k_{n\rbrack}
\cr+{\gamma^{m_1...m_{p-1}}_{\alpha\beta}}
R_{{}m_1...m_p}(k)k_{m_p})
\times
{c\bar{c}}e^{-{3\over2}\phi-{1\over2}{\bar\phi}}
\Sigma_\alpha{\bar{\Sigma}}_\beta{e^{ikX}}(z,\bar{z})
\equiv{V_{RR}^{(-3/2,-1/2)}(k)}}}
where the square brackets imply antisymmetrization over the
space-time indices.
Next, acting
on this expression with $\Gamma(z)$
we get:
\eqn\grav{\eqalign{:\Gamma{\bar\Gamma}{V_{RR}^{(-3/2,-3/2)}(k)}:
={1\over4}{c\bar{c}}e^{-{1\over2}\phi-{1\over2}{\bar\phi}}
\Sigma_\alpha{\bar{\Sigma}}_\beta{e^{ikX}}(z,\bar{z})
\cr\times{2}\Gamma^{nm_1...m_{p-1}}_{\alpha\beta}
k_{m_p}k_{{\lbrack}n}R_{m_1...m_{p-1}\rbrack{m_p}}(k)
\cr\equiv{V_{RR}^{(-1/2,-1/2)}(k)}}}
Comparing this with (1) we deduce the following relations
between the RR field strength $F$, RR gauge potential
$A$ and RR prepotential $R$:
\eqn\grav{\eqalign{F_{nm_1...m_{p-1}}(k)=
k_{m_p}k_{{\lbrack}n}R_{m_1...m_{p-1}\rbrack{m_p}}(k)\cr
A_{m_1...m_{p-1}}(k)=
k_{m_p}R_{m_1...m_{p-1}{m_p}}(k)}}
In other words, the Ramond-Ramond $(p-1)$-form gauge potential
$A$ is given by the divergence  of the
p-form prepotential $R$, while the RR field strength $F$
is given by the Laplacian of $R$.
The p-form prepotential  $R$ may also be interpreted as a parameter for
the Penrose class of solutions to Maxwell's equations
Therefore the RR potential enters the
expression for picture $-3/2,-1/2$
(or equivalently  $-1/2,-3/2$)
RR vertex operator which can now be expressed as
\eqn\grav{\eqalign{{V_{RR}^{(-3/2,-1/2)}(k)}:
={1\over2}{c\bar{c}}e^{-{3\over2}\phi-{1\over2}{\bar\phi}}
\Sigma_\alpha{\bar{\Sigma}}_\beta{e^{ikX}}(z,\bar{z})
(\Gamma^{m_1...m_{p-1}}_{\alpha\beta}A_{m_1...m_{p-1}}(k)
\cr+\Gamma^{m_1...m_{p}n}
k_{{\lbrack}n}R_{m_1...m_{p-1}\rbrack{m_p}}(k))}}
In other words, the RR vertex operator taken at the mixed
 $(-3/2,-1/2)$-picture can be considered as a source
of the RR-charge (shifted by the exterior derivative
of the RR-prepotential, necessary to insure the BRST-invariance).
In this sense the  $(-3/2,-1/2)$-picture vertex operators
are similar to D-branes and should have a
non-perturbative nature, while the structure of their
scattering amplitudes may be expected to
reflect a non-perturbative physics.
In this letter we shall observe and discuss
one particularly interesting example of non-perturbative effect
related to RR scattering amplitudes at non-canonical pictures -
the breaking of the B-field gauge invariance in the low energy
effective action due to the presence of the non-canonical RR states.
Namely, we will show that  terms of the form
$(B{\wedge}F^{(p-2)},F^{(p)})$ appear in the effective Lagrangian
(particularly for $p=4$) due to the interaction
of the B-field with non-canonical RR backgrounds in superstring theory.
Note that, contrary to the
canonical case (related to scattering amplitudes of the B-field
with canonical RR vertices)
 these terms
are not full derivatives (as this is the case for the well-known
CS term $B{\wedge}F^{(4)}{\wedge}F^{(4)}$ coming from M-theory)
and therefore the B-field gauge invariance is broken.
Let us stress that this effect is  non-perturbative
(as the ${V_{RR}^{(-3/2,-1/2)}}$ vertices correspond
to  non-perturbative brane dynamics)
and of course may lead to significant consequences for
issues like non-commutativity occuring in certain
B-field backgrounds. Some of these consequences will be discussed in this
letter.

\centerline{\bf 2. B-field in RR backgrounds and
non-perturbative 2-form state}

Before starting the calculation of scattering amplitudes
revealing the

non-perturbative B-field gauge invariance breaking,
we shall comment on some peculiarities
of the OPE's of spin operators at non-canonical pictures to
clarify the physics behind the gauge invariance breaking.
Consider first the OPE of two canonical Ramond spin operators.
We shall be interested in simple poles of these OPE's
(as for 3-point correlators of primary fields
only these OPE terms are important).
Using the OPE expressions (6) we have:
\eqn\grav{\eqalign{e^{-{1\over2}\phi}\Sigma_\alpha(z)
e^{-{1\over2}\phi}\Sigma_\beta(w)
\sim{1\over{z-w}}e^{-\phi}\psi_m(w)\Gamma^m_{\alpha\beta}+...}}
At the same time,
\eqn\grav{\eqalign{
e^{-{3\over2}\phi}\Sigma_\alpha(z)e^{-{1\over2}\phi}\Sigma_\beta(w)
\sim{1\over{(z-w)^2}}e^{-2\phi}(w)\delta_{\alpha\beta}\cr+
{1\over{z-w}}({1\over2}\partial(e^{-2\phi})(w)\delta_{\alpha\beta}
+{1\over2}e^{-2\phi}\psi_m\psi_n(w)\gamma^{mn}_{\alpha\beta})}}

The OPE (12) differs from (11) substantially.
While the r.h.s. of (11) contains only the usual vector
 field at picture $-1$, the r.h.s.
of the OPE  (12) involving the non-canonical spin field
$e^{-{3\over2}\phi}\Sigma_\alpha$ contains the two-form
term given by  $\Phi_{mn}=e^{-2\phi}\psi_m\psi_n$
The origin of this two-form has been discussed
in ~{\self}. It has been shown that
this corresponds to the  membrane topological charge ~{\azc},
appearing as a two-form central term in picture-changed
space-time SUSY algebra.
This intermediate state may also be interpreted as a
two-form $\Phi$-parameter associated with a choice of regularization
in the worldsheet path integral for a string theory with the B-field
~{\seib,\and}. Indeed, it is easy to see  that for the space-time
conjugate momentum
operator at picture -1 $P_m=\oint{{dz}\over{2i\pi}}e^{-\phi}\psi_m(z)$
one has ${\lbrack}P_m,P_n{\rbrack}=\oint{{dz}\over{2i\pi}}\Phi_{mn}
\sim{\lbrack}\partial_m,\partial_n{\rbrack}$.
At nonzero momenta this two-form
 gives rise to physical
vertex operator given by ${\sim}c{e^{-2\phi}\psi_m\psi_n}e^{ikX}$
which does not correspond to any perturbative open string excitation
(such as a photon) but which
describes the non-perturbative membrane dynamics.
There is no version of this vertex at picture zero;
this operator is BRST-nontrivial if the momentum k
is directed along any of 8 space-time directions
transverse to its indices $m$ and $n$.
Now,
because of the form of the OPE (12)
 these two-form vertices
appear as an intermediate state (both in left and right sectors)
in all amplitudes involving the RR vertices at non-canonical
pictures. The crucial point is that it is the interaction
of this intermediate two-form membrane-like vertex
with the axionic state that makes the amplitude picture-dependent and plays the crucial
role in breaking the gauge invariance of the B-field in the low-energy
effective action. This effect is therefore non-perturbative. In the following section we shall
demonstrate it by direct computation of scattering amplitude.

\centerline{\bf 3. Interaction of the B-field with non-canonical RR
backgrounds}

In this  section we compute the interaction
of the B-field with  2-form RR field strength
and the 4-form RR-prepotential
taking place in the type IIA theory,
 showing that
it gives rise to the anomalous term in
the low-energy effective action.

The relevant correlator to compute is given by

\eqn\lowen{A_{FFB}=<{V_{RR}^{(-3/2,-3/2)}}(p){V_{RR}^{(-1/2,-1/2)}}(k)
V_B^{(0,0)}(q)>}

where $V_B^{(0,0)}(p)$ is the axionic vertex at picture $(0,0)$,
given by
\eqn\lowen{V_B^{(0,0)}(q)={c\bar{c}}
(\partial{X^m}+i(q\psi)\psi_m)({\bar\partial}{X^m}+i(q{\bar\psi}){\bar\psi}_n)
e^{iqX}(z,\bar{z})B_{mn}(q)}
Let us start with computing  the correlator of  RR-vertices with
the purely fermionic part of $V_B$, i.e. the one biquadratic
in $\psi$ and $\bar\psi$.
We have:
\eqn\grav{\eqalign{A_{FFB}^{(1)}(p,k,q)=
-{1\over{2!4!}}<{c\bar{c}}e^{-{3\over2}\phi-{3\over2}{\bar\phi}}
\Sigma_{\alpha_1}{\bar{\Sigma}}_{\beta_1}{e^{ipX}}(z_1,\bar{z_1})
{c\bar{c}}e^{-{3\over2}\phi-{3\over2}{\bar\phi}}
\Sigma_{\alpha_2}{\bar{\Sigma}}_{\beta_2}{e^{ikX}}(z_2,\bar{z_2})
\cr\times
q^sq^t{c\bar{c}}:\psi_s\psi_m{\bar\psi}_t{\bar\psi}_n:(z_3,\bar{z_3})>
{\gamma^{m_1...m_4}_{\alpha_1\beta_1}}
{\gamma^{n_1n_2}_{\alpha_2\beta_2}}
R_{m_1...m_4}(p)F_{n_1n_2}(k)B_{mn}(q)
}}
(the minus sign here is due to the total $i^2$ factor
in the fermionic part of $V_B$)
Computing this correlator  using the OPE  expressions (6) we obtain
\eqn\grav{\eqalign{A_{FFB}^{(1)}(p,k,q)=
-{1\over{{(2!)^3}4!}}
Tr(\gamma^{m_1...m_4}\gamma^{tn}\gamma^{n_1n_2}\gamma^{sm})
q_sq_t
\cr\times
R_{m_1...m_4}(p)F_{n_1n_2}(k)B_{mn}(q)\delta(p+k+q)}}
Straightforward evaluation of the gamma-matrix trace gives:
\eqn\grav{\eqalign{Tr(\gamma^{m_1...m_4}\gamma^{tn}
\gamma^{n_1n_2}\gamma^{sm})\cr=
-(96g^{mn_1}g^{sm_1}g^{n_2m_2}g^{tm_3}g^{nm_4}+perm
(\lbrack{n_1}\leftrightarrow{n_2}\rbrack,
\lbrack{m}\leftrightarrow{s}\rbrack))\cr
-(96g^{mt}g^{sm_1}g^{n_1m_2}g^{n_2m_3}g^{nm_4}+perm
(\lbrack{n}\leftrightarrow{t}\rbrack,
\lbrack{m}\leftrightarrow{s}\rbrack))
\cr
+(96g^{mm_1}g^{sn_1}g^{n_2m_2}g^{tm_3}g^{nm_4}+perm
(\lbrack{n_1}\leftrightarrow{n_2}\rbrack,
\lbrack{m}\leftrightarrow{s}\rbrack))\cr
+(96g^{mm_1}g^{sm_2}g^{n_1m_3}g^{tn_2}g^{nm_4}+perm
(\lbrack{n}\leftrightarrow{t}\rbrack,
\lbrack{m}\leftrightarrow{s}\rbrack))}}
where $g^{mn}$ is Minkowski tensor and
permutations imply antisymmetrizations
over the appropriate indices.
Contracting the obtained expression for the trace with
the space-time fields and using the
on-shell conditions:
\eqn\grav{\eqalign{q^mB_{mn}(q)=0\cr
k^mF_{mn}(k)=0\cr
k_{{\lbrack}m}F_{np\rbrack}(k)=0\cr
q^2=k^2=p^2=0}}
we obtain the result for this part of the amplitude:

\eqn\grav{\eqalign{
A_{FFB}^{(1)}(p,k,q)=
-{{192}\over{{{2!}^3}4!}}
q_sR_{mnsn_1}(p)q^tF_{tn_1}(k)B_{mn}(q)
=-A^{RR}_{mnn_1}(p)p^tF_{tn_1}(k)B_{mn}(q)\cr=
-F_{tmnn_1}(p)F_{tn_1}(k)B_{mn}(q)}}
The next contribution is from the correlator involving the
X-part of the axionic vertex:
\eqn\grav{\eqalign{A_{FFB}^{(2)}(p,k,q)=
{1\over{2!4!}}<{c\bar{c}}e^{-{3\over2}\phi-{3\over2}{\bar\phi}}
\Sigma_{\alpha_1}{\bar{\Sigma}}_{\beta_1}{e^{ipX}}(z_1,\bar{z_1})
{c\bar{c}}e^{-{3\over2}\phi-{3\over2}{\bar\phi}}
\Sigma_{\alpha_2}{\bar{\Sigma}}_{\beta_2}{e^{ikX}}(z_2,\bar{z_2})
\cr\times{c\bar{c}}\partial{X^m}\bar\partial{X^n}e^{iqX}>
\times
{\gamma^{m_1...m_4}_{\alpha_1\beta_1}}
{\gamma^{n_1n_2}_{\alpha_2\beta_2}}
R_{m_1...m_4}(p)F_{n_1n_2}(k)B_{mn}(q)
\cr=k_np_mR_{m_1...m_4}(p)F_{n_1n_2}(k)B_{mn}(q)
Tr(\gamma^{m_1...m_4}\gamma^{n_1n_2})=0
}}
i.e. this contribution iz zero as the  gamma-matrix trace vanishes.
Finally, the cross-term contribution is given by:
\eqn\grav{\eqalign{A_{FFB}^{(3)}(p,k,q)=
{i\over{2!4!}}<{c\bar{c}}e^{-{3\over2}\phi-{3\over2}{\bar\phi}}
\Sigma_{\alpha_1}{\bar{\Sigma}}_{\beta_1}{e^{ipX}}(z_1,\bar{z_1})
{c\bar{c}}e^{-{3\over2}\phi-{3\over2}{\bar\phi}}
\Sigma_{\alpha_2}{\bar{\Sigma}}_{\beta_2}{e^{ikX}}(z_2,\bar{z_2})
\cr\times{c\bar{c}}
(q_s\psi_s\psi_m\bar\partial{X^n}+\partial{X^m}q_s\bar\psi_s\bar\psi_n)
{e^{iqX}}(z_3,\bar{z_3})>\cr\times
{\gamma^{m_1...m_4}_{\alpha_1\beta_1}}
{\gamma^{n_1n_2}_{\alpha_2\beta_2}}
R_{m_1...m_4}(p)F_{n_1n_2}(k)B_{mn}(q)
\cr=-2\times{{i\over{(2!)^24!}}}R_{m_1...m_4}(p)F_{n_1n_2}(k)B_{mn}(q)
q_sq_nTr(\gamma^{m_1...m_4}\gamma^{sm}\gamma^{n_1n_2}\gamma^{m_1...m_4})
}}
Evaluating the
gamma-matrix trace as before
and using the on-shell conditions for the space-time fields
along with momentum conservation we get
\eqn\grav{\eqalign{A_{FFB}^{(3)}(p,k,q)
=-{{8\times{4!}}\over{(2!)^24!}}
q_sR_{smn_1n_2}(p)F_{n_1n_2}(k)p_nB_{mn}(q)
\cr=2A^{RR}_{mn_1n_2}(p)F_{n_1n_2}(k)p_nB_{mn}(q)
=2F_{nmn_1n_2}(p)F_{n_1n_2}(k)B_{mn}(q)
\cr=-2F_{mnn_1n_2}(p)F_{n_1n_2}(k)B_{mn}(q)}}
Physically, the factor of 2 in this contribution is related
to sum of the contributions from the left and the right sectors
(interaction with left and right intermediate two-form states)
Adding all the contributions together,
we get
\eqn\grav{\eqalign{A_{FFB}(p,k,q)\equiv
A_{FFB}^{(1)}(p,k,q)+A_{FFB}^{(2)}(p,k,q)+A_{FFB}^{(3)}(p,k,q)
\cr
=-3F_{mnn_1n_2}(p)F_{n_1n_2}(k)B_{mn}(q)}}
This concludes our calculation of the 3-point correlator.
It corresponds to the term
\eqn\lowen{S_{BFF}\sim\int{d^{10}}{x}F_{mnn_1n_2}F_{n_1n_2}B_{mn}}
in the low-energy effective action, apparently
breaking the B-field gauge symmetry.
As we have already remarked above, the mechanism of this symmetry
breaking
originates from the interaction of the B-field vertex
with
the membrane-like intermediate states (left and right), described by two-form
vertex operators at the picture $-2$,
i.e. this effect is  non-perturbative,
even though technically it involves only the
perturbative string amplitudes.
In the next section we shall discuss the relevance of this
result to the space-time non-commutativity problem.

\centerline{\bf 4. Gauge symmetry breaking and non-commutativity in RR
backgrounds}

The phenomenon of the gauge invariance breaking, caused by
the non-perturbative interaction
of the B-field with non-canonical RR-states, particularly raises
questions of how consistent are the models of space-time noncommutativity
based on open strings in a constant B-field background
In these models the presence of the B-field modifies
the two-point propagator $<X^m(z)X^n(w)>$ on the worldsheet boundary.
As a result the modified propagator acquires  the antisymmetric part
giving rise to non-commutativity of space-time coordinates after the
regularization at coincident points z and w ~{\seiberg, \shahin}
 The non-commutativity parameter, $\theta_{mn}$,
is then given by the function of the $B_{mn}$ axionic field,
originating from a rank 2 antisymmetric massless mode
of a closed string.
In these models the two-form $B$-field is treated as a static
background, without any regard to a closed string.
In the perturbative string-theoretic framework such a consideration
is valid and non-contradictory since perturbatively the
constant $B$-field background plays no role in closed string dynamics.
This is because perturbatively the B-field is a gauge field,
entering the low-energy effective action through its
3-form field strength $H=dB$. The only exception is
the CS term $B\wedge{F^{(4)}}\wedge{F^{(4)}}$
where $F^{(4)}$ is the RR 4-form field strength.
This topological term, originating from M-theory,
 is gauge-invariant and does not affect the equations of motion,
playing the role analogous to the $\theta$-term.
For constant $B_{mn}$-field the B-field strength is
zero and it can be gauged away by suitable transformation.
The situation changes, however,
if one takes into account the non-perturbative interaction
of the B-field with the RR-sector (taking place through the
intermediate NS 2-form state $e^{-2\phi}\psi_m\psi_n{e^{ikX}}$)
Due to the non-vanishing s-matrix element of the B-field with
two RR-states. As a result, the gauge symmetry is broken and the
B-field  background becomes  $dynamical$ and gets
renormalized by the Ramond-Ramond space-time fields.
Due to this renormalization, the space-time profile
of the B-field is no longer arbitrary but is
ralated to the profile of the RR-fields so as to satisfy the
condition of the worldsheet conformal invariance.
To see the relation consider the
worldsheet RG flow equations involving the B-field
and the RR-states, taking into account their non-perturbative
interaction (13).Using the result of our calculation
of the 3-point correlator (23)
it is easy to write down equations for the beta-functions:
\eqn\grav{\eqalign{
\beta_B\equiv
{{dB_{mn}}\over{d(log\Lambda)}}\sim
-k^2{B_{mn}}-3{F_{mnpq}}^{RR(4)}{F_{pq}}^{RR(2)}+...\cr
\beta_{F^{(2)}}\equiv{{d{F_{pq}}^{RR(2)}}\over{d(log\Lambda)}}\sim
-k^2{F_{pq}}^{RR(2)}-3{F_{mnpq}}^{RR(4)}{B_{mn}}+...\cr
\beta_{F^{(4)}}\equiv{{d{F_{mnpq}}^{RR(4)}}\over{d(log\Lambda)}}\sim
-k^2{F_{mnpq}}^{RR(4)}-3{F_{{\lbrack}pq}}^{RR(2)}{B_{mn\rbrack}}}}
Upon the Fourier transform, the conformal invariance
condition $\beta_B=0$ particularly implies the equation of motion for
the B-field:
\eqn\grav{\eqalign{\nabla^2{B_{mn}}=-3{F_{mnpq}}^{RR(4)}{F_{pq}}^{RR(2)}}}
reflecting the non-perturbative breaking of the gauge symmetry.
Thus we see that for general RR backgrounds $F^{RR}\neq{0}$
constant B-field is not a solution of this equation.
In other words, standard non-commutativity models
based on open superstring theory in a constant  B-field
are not conformally invariant on the worldsheet
(on the non-perturbative level) in the presence
of the RR-fields.
One way to cure this problem and to restore the conformal
symmetry along the gauge symmetry is to introduce appropriate
D-brane configuration which would screen the RR-charges.
In terms of the effective action this would correspond
to shifting the B-field as $B\rightarrow{B+dA}$
where $A$ is the D-brane's U(1) field.
As a result, the  terms with the B-field
will be transformed into those with the
 the $B_{mn}+F_{mn}$ - type structure;
the terms that were breaking the
B-field gauge symmetry before the introduction of D-branes
would evolve into those of the type
$(({C}^{RR(p)}+{C}^{RR(p-2)}\wedge(B+dA))^2$
 which are gauge invariant under the appropriate combination of
the U(1) and B field gauge transformations.
In other words, the gauge symmetry will be restored by
the D-brane's U(1) field.
Therefore the introduction of D-brane backgrounds is indispensable
to preserve the B-field gauge symmetry on the non-perturbative level,
as well as to consistently formulate non-commutative theories
in Ramond-Ramond backgrounds, based on a constant
B-field.
\centerline{\bf 5. Conclusions}
In this letter we have considered the Ramond-Ramond vertex operators
at non-canonical pictures, showing that they may be considered as sources
of the RR gauge potential. In this way these operators
have a non-perturbative character (similarly to D-branes)
and correlation functions involving these operators
contain essential information about the non-perturbative physics
of strings.
We also have discussed one particularly interesting
non-perturbative effect related to these vertices
- the breaking of the B-field gauge symmetry.
The only way to restore this gauge symmetry is to introduce D-branes
so that the presence of the D-brane  U(1) field
compensates for the B-field gauge non-invariance.
Therefore the presence of the
non-canonical vertex operators in superstring
spectrum automatically entails the
introduction of D-branes to insure the gauge symmetry.
The gauge symmetry restoration may also be understood in
another way.
Namely, the dynamics of various D-branes can be represented
in terms of the NS and NS-NS brane-like vertex
operators ~{\self} similar to those appearing as
intermediate extra states in the amplitude (23).
As we have seen, these intermediate extra states are in fact those
leading to the gauge symmetry violation.
Therefore introducing D-brane backgrounds (described by the brane-like
states)
is equivalent to screening these intermediate poles.
Such a screening insures that the gauge symmetry associated
with the B-field is restored.

\centerline{\bf Acknowledgements}

We would like to thank A. Kobakhidze,
A.M. Polyakov and M. Sheikh-Jabbari for useful
discussions and comments.
The authors gratefully
acknowledge the financial support of the Academy of Finland
under
the Project no. 54023
D.P. also expresses his acknowledgements to the RFFI public grant
RFBR 01-02-17682

\listrefs
\end